\documentclass[doublecol]{epl2} 

\usepackage{hyperref}
\usepackage{graphicx}
\usepackage{multirow,booktabs}

\usepackage{amsmath,amssymb,latexsym,MnSymbol}

\title{Extensive Characterization of Seismic Laws in Acoustic Emissions of Crumpled Plastic Sheets}

\shorttitle{Seismic Laws in Acoustic Emissions of Crumpled Plastic Sheets} 

\author{Leandro S. Costa\inst{1}, Ervin K. Lenzi\inst{2}, Renio S. Mendes\inst{1} \and Haroldo V. Ribeiro~\inst{1,\hspace{-0.2cm}}\thanks{\email{hvr@dfi.uem.br}}}
\shortauthor{L. S. Costa \etal}

\institute{                    
  \inst{1} Departamento de F\'isica, Universidade Estadual de Maring\'a, Maring\'a, PR 87020-900, Brazil\\
  \inst{2}Departamento de F\'isica, Universidade Estadual de Ponta Grossa, PR 84030-900, Brazil
}


\abstract
{
Statistical similarities between earthquakes and other systems that emit cracking noises have been explored in diverse contexts, ranging from materials science to financial and social systems. Such analogies give promise of a unified and universal theory for describing the complex responses of those systems. There are, however, very few attempts to simultaneously characterize the most fundamental seismic laws in such systems. Here we present a complete description of the Gutenberg-Richter law, the recurrence times, Omori's law, the productivity law, and B{\aa}th's law for the acoustic emissions that happen in the relaxation process of uncrumpling thin plastic sheets. Our results show that these laws also appear in this phenomenon, but (for most cases) with different parameters from those reported for earthquakes and fracture experiments. This study thus contributes to elucidate the parallel between seismic laws and cracking noises in uncrumpling processes, revealing striking qualitative similarities but also showing that these processes display unique features.
}

\begin{document}

\maketitle

\section{Introduction}
The investigation of earth-related systems has always been present in the physicists' agenda. A remarkable example is the case of earthquakes where several results have been obtained. Actually, the most fundamental seismic laws are intimately connected to the concepts of scaling and universality: $i)$ the Gutenberg-Richter law~\cite{Utsu_intro,Godano_intro} states that the radiated energy $E$ of earthquakes is distributed according to a power law, that is, $P(E)\sim E^{-\beta^\prime}$; $ii)$ the time intervals $\tau$ (recurrence or waiting times) between earthquakes with energy above a lower bound $E_{\text{min}}$ is self-similar over $E_{\text{min}}$ (and over different regions) and its probability distributions can be adjusted by a unique function after rescaling the empirical distributions by the mean seismic activity rate (the so-called unified scaling law~\cite{Bak_intro,Corral_intro0,Corral_intro,Saichev_intro,Davidsen_intro,Touati}); $iii)$ Omori's law~\cite{Utsu_intro2} establishes that the number of aftershocks per unit of time, $R_{\text{a}}(t_{\text{ms}})$, decays as a power-law function of the elapsed time since the mainshock, $t_{\text{ms}}$, that is, $R_{\text{a}}(t_{\text{ms}})\sim t_{\text{ms}}^{-p}$; $iv)$ the productivity law~\cite{Helmstetter_intro} implies that the number of aftershocks, $N_{\text{a}}(E_{\text{ms}})$, triggered by a mainshock of energy $E_{\text{ms}}$ is related to $E_{\text{ms}}$ via  $N_{\text{a}}(E_{\text{ms}})\sim E_{ms}^\alpha$; and finally, $v)$ B{\aa}th's law~\cite{Bath,Helmstetter_intro2} states that the relative difference in energy magnitude (that is, $\log E$) between the mainshock and its largest aftershock is (on average) close to $1.2$, regardless of the mainshock magnitude.

In addition to being observed for earthquakes, some of these laws have been individually reported in very diverse contexts, and particularly in the context of fractures of materials~\cite{Hirata,Diodati,Weiss,Davidsen,Kun,Kun2,Niccolini,Niccolini2,Niccolini3,Salje,Nataf,Nataf2,Baro,Makinen,Ribeiro,Tsai2}, where they find a broad range of applications in engineering and also serve as laboratory experiments for modeling earthquakes. All these phenomena have the common feature of producing impulsive and discrete events (crackling noises) of extremely varied sizes~\cite{Sethna2}. A typical everyday example of crackling noise is observed in the process of crumpling/uncrumpling thin sheets of plastic such as those in plastic bags. Crumpling is a complex process that has attracted the attention of scholars working in several disciplines, from materials science to math~\cite{Witten,Marder}. In the context of acoustic emissions, plastic sheet crumpling was (to the best of our knowledge) first investigated by Kramer and Lobkovsky~\cite{Kramer}, where they found that the energies of the acoustic events have time correlations decaying according to a stretched exponential and are distributed as a power law (see also~\cite{Mendes}). Power-law distributions in energy were also found by Houle and Sethna~\cite{Houle} for paper crumpling, and by Salminen~\textit{et al.}~\cite{Salminen} and Koivisto~\textit{et al.}~\cite{Koivisto} for paper peeling. Koivisto~\textit{et al.} also reported statistics for waiting time between events, where approximated power-law distributions were also found.

Despite the interest in studying cracking noises in crumpling/uncrumpling processes, investigations have hitherto been mainly focused on characterizing energy distributions associated to the acoustic emissions, that is, the analogous of the Gutenberg-Richter law for earthquakes. We still lack a complete parallel (such as those reported in Refs.~\cite{Baro,Ribeiro} for material fractures) with the other four fundamental seismic laws previously mentioned. Here we fill this hiatus by presenting an extensive parallel between the acoustic emissions of plastic sheets and earthquakes laws. Specifically, we characterize: $i)$ the energy distribution (the Gutenberg-Richter law), $ii)$ the recurrence times between events with energy larger than a lower bound (the unified scaling law) and $iii)$ the sequence of aftershocks and foreshocks events (Omori's, B{\aa}th's, and the productivity law) for the cracking noise that plastic sheets emit after crumpled. In the following, we describe the experiments used to obtain the acoustic emissions, the data analysis, and some concluding remarks.

\section{Experiments and Data Acquisition}
In the experiments, plastic sheets (made of biaxially oriented polypropylene) with thickness of $21.0\pm0.1~\mu$m, density of $1.1\pm0.1$~g/cm$^3$, and area of $6.00\pm0.01$~m$^2$ ($1.0$~m~$\times$~$6.0$~m) are crumpled to form a compact ball (by hands and as small as we could make it -- radius around 8~cm). This configuration is kept for about 5~s. The ball is then released and starts to unfold and produce acoustic emissions, which are recorded by a microphone (Shure Microflex MX202W/N) positioned $\sim$$30$~cm from the center of the crumpled sheet, with a sampling rate of 48~kHz. Plastic sheets of this size emit sound for about 40~min and partially restore their original form (around 1/3 of their initial area) after the end the acoustic emissions. 

\begin{figure}[!ht]
\centering
\includegraphics[scale=0.22]{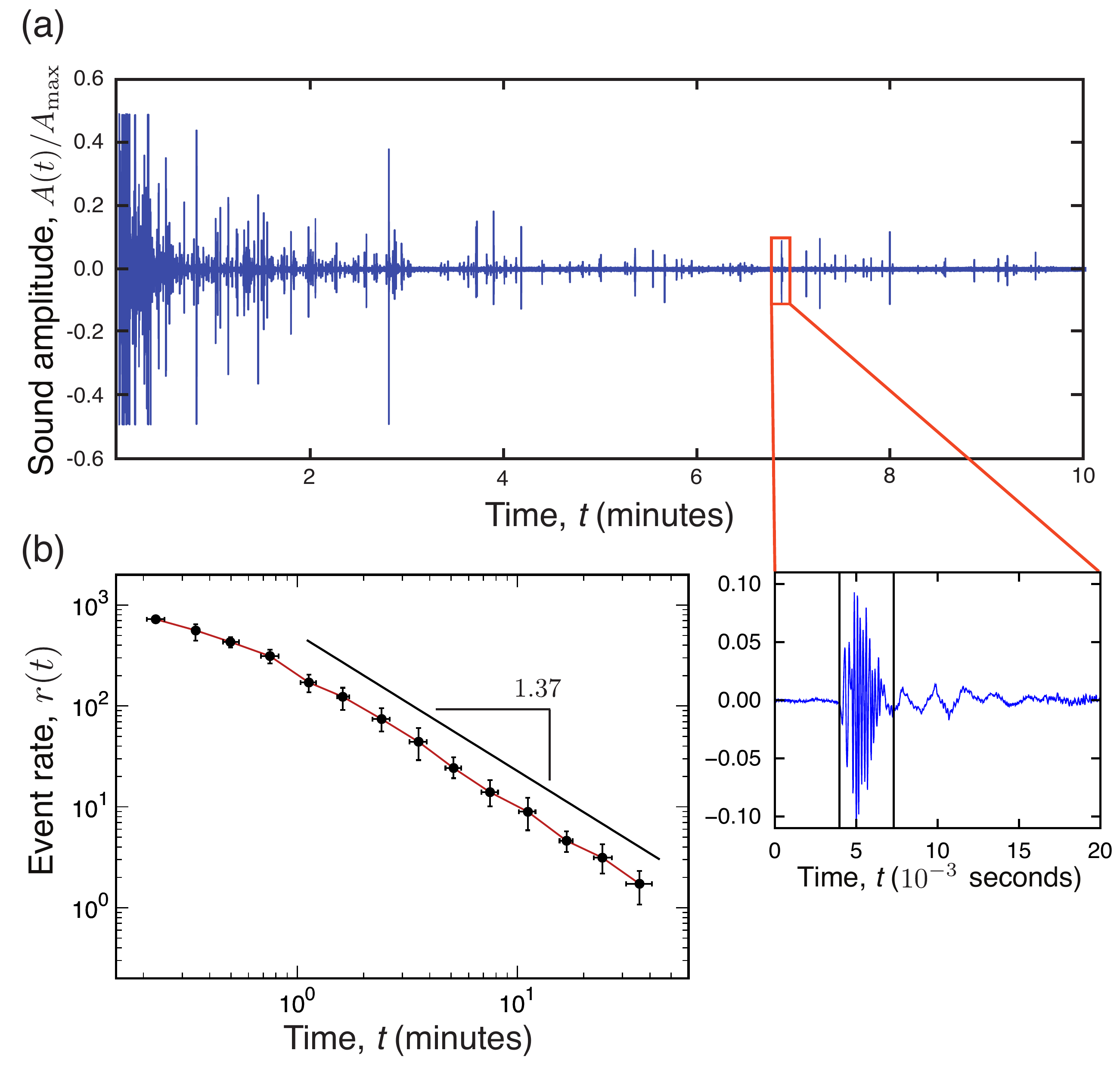}
\caption{Typical behavior of acoustic emissions of crumpled plastic sheets. (a) Normalized sound amplitudes $A(t)/A_{\text{max}}$ obtained in a particular experiment ($A_{\text{max}}=2^{15}-1$ is the saturation limit). The emissions occur as discrete events (the inset shows one event) of varied sizes. (b) Evolution of the rate of activity $r(t)$, that is, the number of events per minute for all samples. The dots are window average over all samples and error bars stand for 95\% bootstrap confidence intervals. Notice that $r(t)$ is non-stationary and decays approximately as a power law ($r(t) \sim t^{-1.37}$, for $t > 1$~min).
}
\label{fig:1}
\end{figure}

Figure~\ref{fig:1}(a) shows the sound amplitudes $A(t)$ (rescaled the saturation limit $A_{\text{max}}=2^{15}-1$) recorded during a particular experiment. In all experiments, sound card and preamplifier gains were constant and the same. This values were adjusted in order to not saturate the microphone and keep the maximum sound amplitude during the recordings in around $A_{\text{max}}/2$. An event $i$ is identified by a starting time $t_{\text{ini}_{i}}$ and ending time $t_{\text{end}_{i}}$. These times are obtained by analyzing the normalized sound intensity $I(t) = {A(t)^2}/{\text{max} [A(t)^2]}$. Specifically, $t_{\text{ini}_{i}}$ is defined as the time for which $I(t)$ initially exceeds a threshold $I_{\text{min}}$, and $t_{\text{end}_{i}}$ is defined as the time for which $I(t)$ stays below $I_{\text{min}}$ for more than $\Delta t$ units of time. The energy of an event $i$ is calculated via $E_{i} = \int_{t_{\text{ini}_{i}}}^{t_{\text{end}_{i}}} I(t)\, dt$ (integrate in time-steps of minutes) and the location time associated to this event is $t_{i} = (t_{\text{end}_{i}} + t_{\text{ini}_{i}})/2$. 

Our results are based on experiments with 16 samples and for $I_{\text{min}}=5\times10^{-4}$ and $\Delta t=0.025$~s. We empirically verified that these values correctly identified the events after dropping the initial 30~s of the recordings (there are several overlapping events at the beginning of the recordings, a fact that can be verified by studying the duration of the events). We have further verified that very similar results are obtained for $I_{\text{min}}$ from $10^{-4}$ to $2 \times 10^{-3}$ and $\Delta t$ from $0.025$~s to $0.2$~s. We further observe that the acoustic emissions are not stationary (as one can observe directly from Fig.~\ref{fig:1}(a)); in fact, the rate of activity $r(t)$, defined as the number of events per minute, decays approximately as a power-law function of time $t$, $r(t)\sim t^{-\eta}$, with $\eta=1.37\pm0.02$ as depicted in Fig.~\ref{fig:1}(b). 

\begin{figure*}[!ht]
\centering
\includegraphics[scale=0.22]{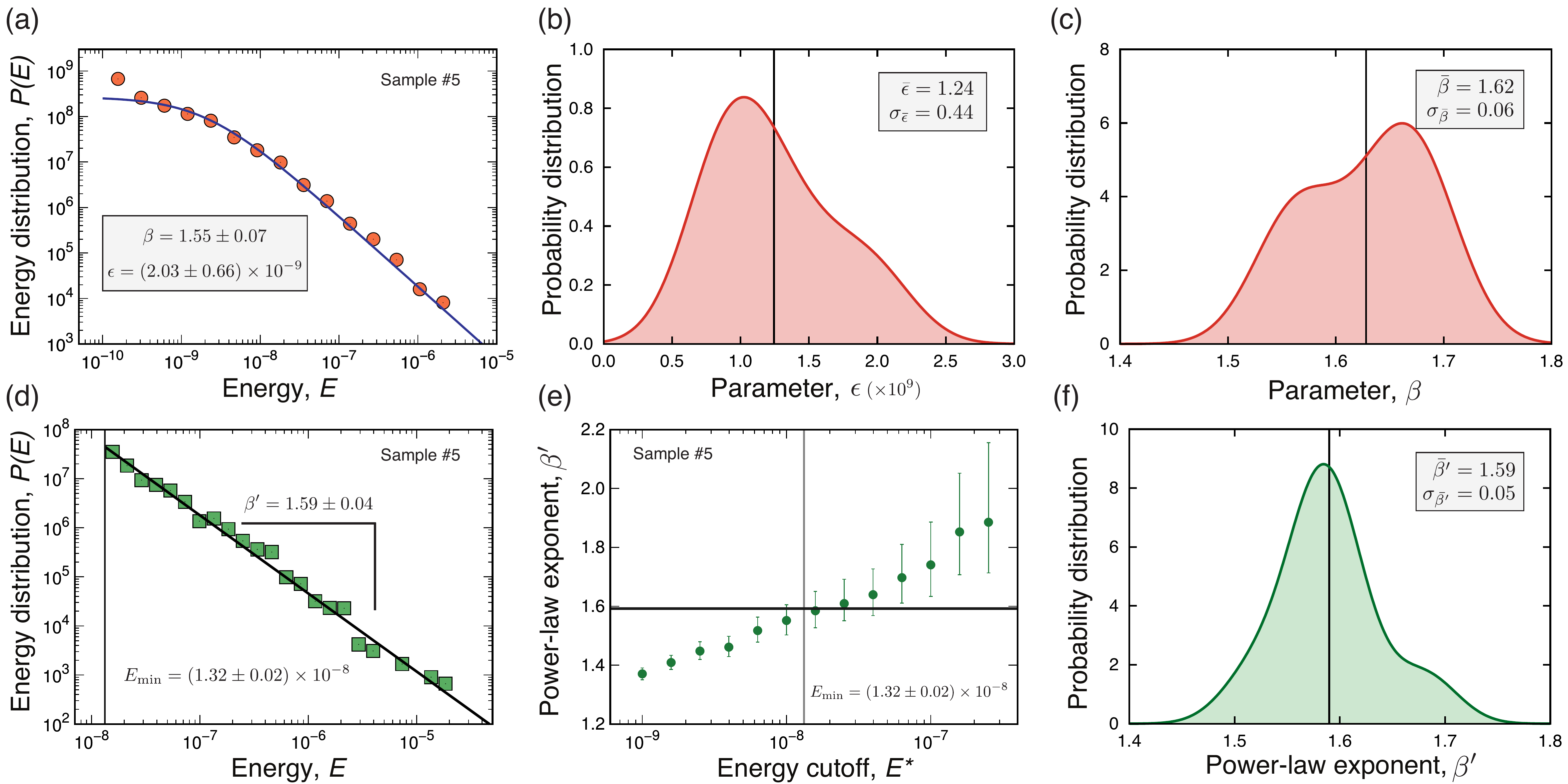}
\caption{The Gutenberg-Richter law. (a) The circles show the energy distribution $P(E)$ (in units of inverse of sound intensity times minutes) calculated with data from one experiment (sample \#5). The continuous line is the adjusted Zipf-Mandelbrot distribution, $P(E)=[(\beta -1) \varepsilon ^{\beta -1}]/[E+\varepsilon]^\beta$, with $\beta=1.55\pm0.07$ and $\varepsilon=(2.0\pm0.7)\times10^{-9}$ obtained via maximum likelihood method. The $p$-value of the Kolmogorov-Smirnov test is $0.19$, indicating the Zipf-Mandelbrot hypothesis cannot be rejected from data. Very similar agreements are obtained for all samples, in particular, all $p$-values are larger than $0.05$. Probability distributions of the values of (b) $\varepsilon$ and (c) $\beta$ over all samples. The average values (indicated by vertical lines) and the standard deviations of these parameters are shown in the plots. Panel (d) shows the tail of the energy distribution ($E>E^*$, with $E^*=E_{\text{min}} = (1.32\pm0.02)\times 10^{-8}$), that can be approximated by a ``pure'' power-law, $P(E)\propto E^{-\beta'}$, with $\beta'=1.59\pm0.04$, and panel (e) illustrates how the maximum likelihood estimation of $\beta'$ depends on the lower energy cutoff $E^*$. The horizontal line indicates the value of $\beta'$ that minimizes the Kolmogorov-Smirnov statistic. (f) Probability distribution of $\beta'$ over all samples. The average value of $\beta'$ (indicated by a vertical line) and its standard deviation are shown in the plot. The probability distributions for the parameters $\varepsilon$, $\beta$ and $\beta'$ are kernel density estimation with bandwidth given by Scott's rule.
}
\label{fig:2}
\end{figure*}

\section{Data Analysis}
\textit{Gutenberg-Richter Law.} We start by calculating the energy distribution. This aspect was also investigated by Kramer and Lobkovsky~\cite{Kramer} and by Mendes~\textit{et al.}~\cite{Mendes} for plastic sheets, where power-law distributions were reported for the energies, compatible with the Gutenberg-Richter law. On the other hand, Tsai~\textit{et al.}~\cite{Tsai} have recently argued that the Zipf-Mandelbrot distribution is a better fit for the energy distribution of single crumpled sheets composed of aluminum, high-density polyethylene, or A4 copy paper. For our data, we compare four candidate distributions: log-normal, power law, Weibull, and Zipf-Mandelbrot via Akaike information criterion. As in Tsai~\textit{et al.}~\cite{Tsai}, the Zipf-Mandelbrot distribution, $P(E)\propto (E+\varepsilon)^{-\beta}$ ($\varepsilon$ and $\beta$ are the model parameters obtained via maximum likelihood), is the best model for all samples. The same conclusion is obtained by likelihood-ratio tests and the Kolmogorov-Smirnov test cannot reject the Zipf-Mandelbrot hypothesis at a confidence level of 99\%. Figure~\ref{fig:2}(a) shows one of the energy distributions and the Zipf-Mandelbrot fit, where a good agreement is observed. Figures~\ref{fig:2}(b)~and~\ref{fig:2}(c) show the distributions of the fitting parameters ($\beta$ and $\varepsilon$) over all samples. We observe that $\beta$ is characterized by an average value of $\bar{\beta}=1.62$ with standard deviation of $\sigma_{\bar{\beta}}=0.06$, and that $\varepsilon$ has an average of $\bar{\varepsilon}=1.24\times 10^{-9}$ and standard deviation of $\sigma_{\bar{\varepsilon}}=0.44\times 10^{-9}$. As it also discussed in Tsai~\textit{et al.}~\cite{Tsai}, deviations from a pure power-law distribution may be partially explained by the attenuation of sound intensity caused by the multiple layers that a crumpled ball contains. In our case, sound attenuation may be further caused by the increasing distance of ends the sheets from the microphone as the experiment advances (usually varying from $0.3$ to $1$ meter). Under this hypothesis, one may expect the energy distribution to be non-stationary; however, we find no clues of this behavior in the energy time series. In particular, we have observed that the form of the energy distribution practically does not change along the experiment and that the $\beta$ is also constant over time.

Despite the Zipf-Mandelbrot model being a better fit for the energy distributions, a more direct comparison with the Gutenberg-Richter law is obtained by fitting a power law to the tails of the energy distributions, that is, considering only energies larger than $E^*$. The best value for $E^*$ can be obtained through the Clauset~\textit{et al.}~\cite{Clauset} approach for fitting power laws. This method basically consists of adjusting (via method of maximum likelihood) a power-law distribution, $P(E)\propto E^{-\beta'}$ for $E>E^*$ ($\beta'$ is the power-law exponent), for a range of $E^*$ values and choosing the best $E^*=E_{\text{min}}$ as the one that minimizes the Kolmogorov--Smirnov statistic. Figure~\ref{fig:2}(d) shows the tail of the energy distribution estimated by the previous approach, where $E_{\text{min}}=(1.32\pm0.05)\times 10^{-8}$ and $\beta'=1.59\pm0.04$, and Fig.~\ref{fig:2}(e) shows $\beta'$ versus $E^*$, both for sample \#5. We notice that $\beta'$ has a systematic increasing trend with the values of $E^*$ (similar behavior is observed for all samples). This behavior agrees with the Zipf-Mandelbrot hypothesis (since it bends downward for small $E$), but differs from the results reported for earthquakes and fracture of materials. For earthquakes, $\beta'$ versus $E^*$ usually displays an initial increasing trend followed by a decreasing trend~\cite{Godano_intro}; whereas an approximated plateau is observed for fracture of materials~\cite{Baro,Ribeiro}. Figure~\ref{fig:2}(f) shows the distribution of $\beta'$ (for $E^*=E_{\text{min}}$) over all samples, where we observe that $\beta'$ has an average value of $\bar{\beta'}=1.59$ and a standard deviation of $\sigma_{\bar{\beta'}}=0.04$ (values that are similar to the ones reported for $\beta$). This average value is close to the ones reported for earthquakes ($\beta'\approx1.67$~\cite{Utsu_intro,Godano_intro}) and greater than those observed for fracture of materials (\textit{e.g.} $\beta'\approx1.40$ for Vycor~\cite{Baro} and wood~\cite{Makinen}, $\beta'\approx1.45$ for bamboo, and $\beta'\approx1.30$ for charcoal~\cite{Ribeiro}).

\begin{figure}[!ht]
\centering
\includegraphics[scale=0.22]{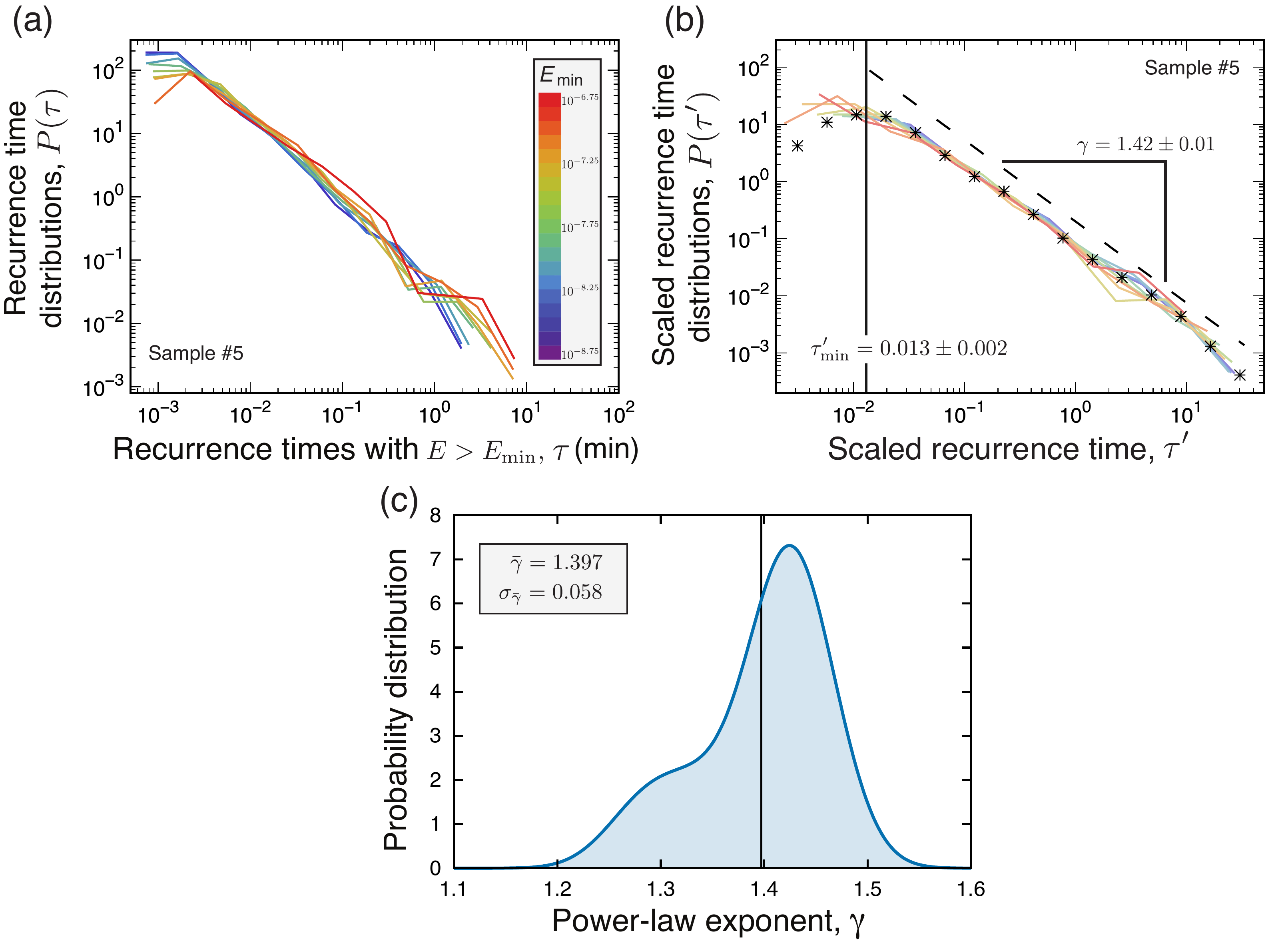}
\caption{Recurrence times between events. (a) Each curve shows the probability distribution $P(\tau)$ for the recurrence times $\tau$ (in minutes) with $E>E_{\text{min}}$ for a particular sample (sample \#5). The color code refers to the values of $E_{\text{min}}$. Panel (b) shows the same distributions when considering the rescaled recurrence time, $\tau^\prime=\langle r\rangle\, \tau $, with $\langle r\rangle$ being the average rate activity (number of events per unit of time). The asterisk marks show the distribution after aggregating the all values of $\tau^\prime$ for different $E_{\text{min}}$. The vertical line indicates the beginning of the power-law regime ($\tau^\prime_{\text{min}}=0.013\pm0.002$) and the dashed line shows the adjusted power law with $\gamma=1.42\pm0.01$. Similar plots are observed for all samples. (c) Probability distribution (kernel density estimation with bandwidth given by Scott's rule) of $\gamma$ over all samples. The average value of $\gamma$ (indicated by a vertical line) and its standard deviation are shown in the plot.
}
\label{fig:3}
\end{figure}

\textit{Recurrence times between events.} A recurrence time $\tau$ is defined as the time interval separating two acoustic events with energy larger than $E_\text{min}$. For earthquakes, Corral~\cite{Corral_intro0,Corral_intro} proposed that the distributions of $\tau$ (for different values of $E_\text{min}$) collapse onto single curve (that is, they are self-similar) after multiplying the recurrence times by the rate of seismic activity $r$ (number of events per unit of time), a fact that was later explained by Saichev and Sornette~\cite{Saichev_intro} as an emergent property of aftershock superposition (see also~\cite{Touati}). The functional form of $P(\tau)$ is usually approximated by a power law when $r$ is time dependent, and by a gamma distribution when $r$ is nearly stationary. For instance, $P(\tau)$ was found to follow power laws for the cracking noise of a porous material under compression~\cite{Baro}, whereas a gamma distribution describes the recurrence times between acoustic events in charcoal samples under irregularly distributed internal stresses~\cite{Ribeiro}. In paper peeling experiments~\cite{Koivisto}, an analysis of the recurrence time distribution with $E_\text{min}=0$ found an approximate power-law regime that is dependent on the driving forces. 

For our experiments, Fig.~\ref{fig:3}(a) shows the distributions of recurrence times $\tau$ for a sample and for several values of $E_\text{min}$. Figure~\ref{fig:3}(b) shows the same distributions after rescaling the recurrence times $\tau$ by the average rate activity $\langle r\rangle$, where a good collapse is observed. We further note that the collapsed distributions seem to decay as a power-law function, that is, $P(\tau^\prime)\sim (\tau^\prime)^{-\gamma}$, with $\tau^\prime=\langle r\rangle\, \tau $ being the rescaled recurrence times and $\gamma$ the power-law exponent. In order to verify the power-law hypothesis, we have aggregated the values of $\tau^\prime$ for all values $E_\text{min}$ and applied the procedure of Clauset~\textit{et~al.}~\cite{Clauset} for each sample. The $p$-values of the Kolmogorov--Smirnov test are all larger than $0.05$, indicating that the power-law hypothesis cannot be rejected for any of the samples. The value of $\tau^\prime_{\text{min}}$ (where the power-law regime begins) and the power-law exponent $\gamma$ are also shown in Fig.~\ref{fig:3}(b) for a particular sample. Very similar behaviors are observed for all samples, and the distribution of $\gamma$ over all samples is shown in Fig.~\ref{fig:3}(c). The average of $\gamma$ is $\bar{\gamma}=1.39$ and its standard deviation is $\sigma_{\bar{\gamma}}=0.06$. {This behavior differs from earthquakes where a crossover between two power-laws is usually observed~\cite{Corral_intro0,Corral_intro}. As discussed by Touati~\textit{et~al.}~\cite{Touati}, this crossover results from the mixture of correlated aftershocks (at short times) and independent events (at large times) related to the spatial heterogeneity of earthquake occurrence, an situation that appears only when recurrence times are evaluated in different spatial windows. Another mechanism that yields two power-law regimes is the temporal evolution in the activity rate. In this case, an initial power-law decay with exponent close $1$ (due the Omori's law) followed by a second power-law decay with exponent $2+1/\eta$ is expected when the activity rate is a power law function of time ($r(t)\sim t^{-\eta}$)~\cite{Baro}. For our case, one would expect a second exponent around $2.7$, a much larger value than the observed ones. On the other hand, as discussed by Saichev and Sornette~\cite{Saichev_intro}, the power-law form of $P(\tau^\prime)$ and its scaling behavior is only an approximation of a more complex behavior (a mixture of power-laws and exponential decays) directly related to the Gutenberg-Richter and Omori laws.}

\begin{figure}[!ht]
\centering
\includegraphics[scale=0.22]{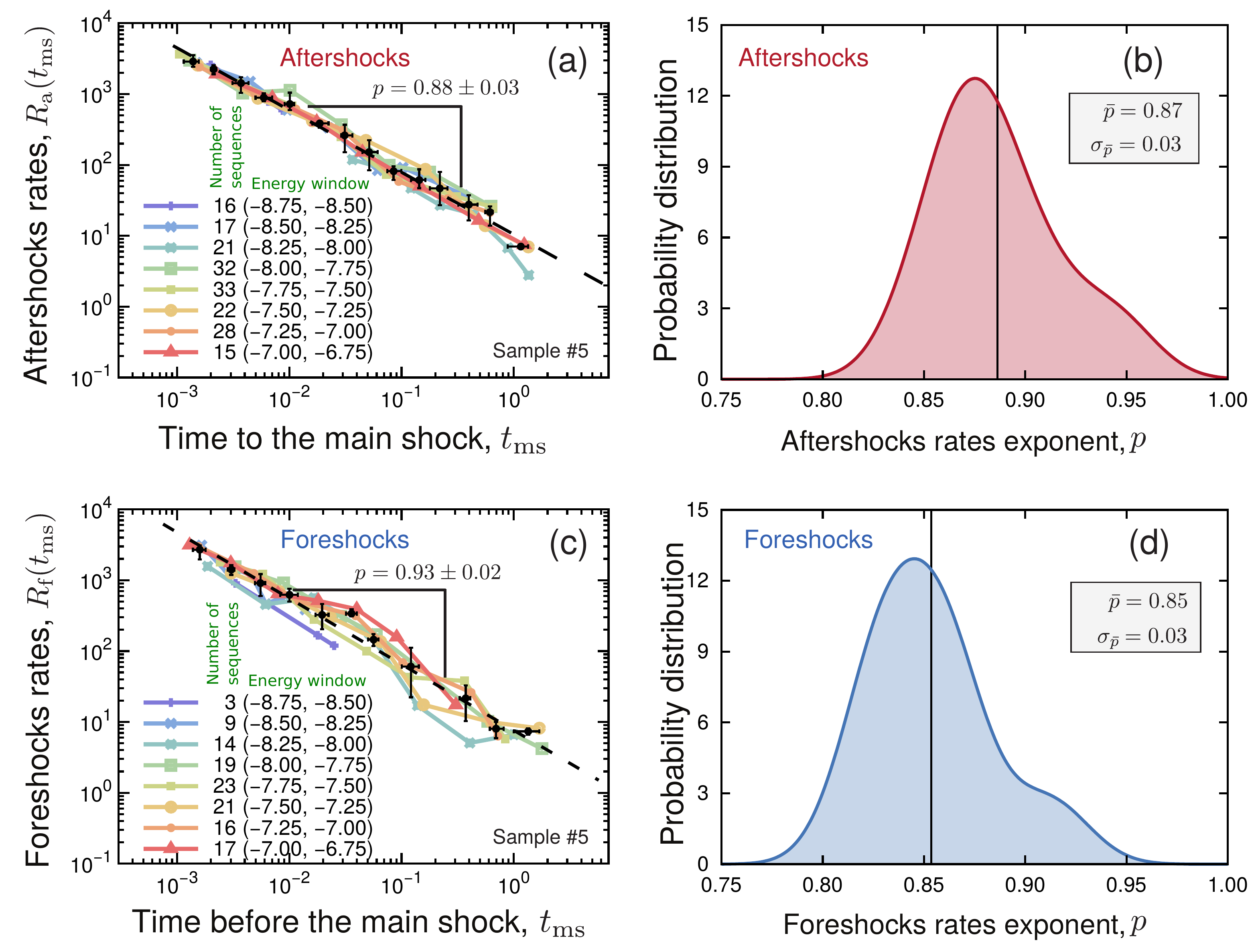}
\caption{The Omori's law for aftershocks and foreshocks. (a) Each colorful curve (and mark) shows the number of aftershocks per unit of time (aftershock rates), $R_{\text{a}}(t_{\text{ms}})$, as a function of the time to mainshock, $t_{\text{ms}}$ (in minutes), for a given energy window and for one sample (sample \#5). The number of sequences in each energy window (and the window as well) is shown in the plot. The black dots are window average values over all curves and the error bars stand for 95\% bootstrap confidence intervals. The dashed line is a power law adjusted (via ordinary least squares method) to the average Omori decay, where the power-law exponent $p=0.88\pm0.03$ is also shown in the plot. (b) Probability distribution (kernel density estimation with bandwidth given by Scott's rule) of the power-law exponents $p$ over all samples; the average value of $p$ (indicated by a vertical line) and its standard deviation are shown in the plot. Panels (c) and (d) show the same analysis for the foreshock rates, $R_{\text{f}}(t_{\text{ms}})$, where a quite similar behavior is observed.}
\label{fig:4}
\end{figure}

\textit{Omori's Law.} Our next step has been to define the mainshock events and the sequence of aftershocks for studying the Omori's law. {Similarly to other acoustic experiments on material fractures~\cite{Baro,Ribeiro}, we consider as mainshocks all events with energy $E_{\text{ms}}$ in the range $[10^{\epsilon_j},10^{\epsilon_{j+1}}]$ (with $\epsilon_j=-8.75,-8.50,-8.25,\dots,-6.75$) and sequences of aftershocks are defined as the events that follow a mainshock until an event with an energy larger than the energy of mainshock is found.} By using these definitions, we calculate the average aftershock rates $R_\text{a}(t_{\text{ms}})$ (that is, the number of aftershock per unit of time) as a function of the elapsed time since the mainshock ($t_{\text{ms}}$) for each energy window. Figure~\ref{fig:4}(a) shows these rates for each energy window for a given sample. These results reveal a clear tendency to follow the Omori's law, that is, the aftershock rates show a robust power-law decay, $R_\text{a}(t_{\text{ms}})\sim t_{\text{ms}}^{-p}$, that holds for about four decades. The power-law exponent estimated from this sample is $p=0.88\pm0.03$. The distribution of $p$ over all samples is shown in Fig.~\ref{fig:4}(b), where we find that $p$ has an average of $\bar{p}=0.87$ and a standard deviation of $\bar{p}=0.03$. These values are slightly smaller than those reported for earthquakes ($p\in[0.9,1.5]$~\cite{Utsu_intro2}) but close to those reported for fracture experiments~\cite{Baro,Hirata}. The same picture is observed for foreshock rates, with values of $p$ indistinguishable from the aftershock analysis, as shown in Figs.~\ref{fig:4}(c)~and~\ref{fig:4}(d).

\begin{figure}[!t]
\centering
\includegraphics[scale=0.22]{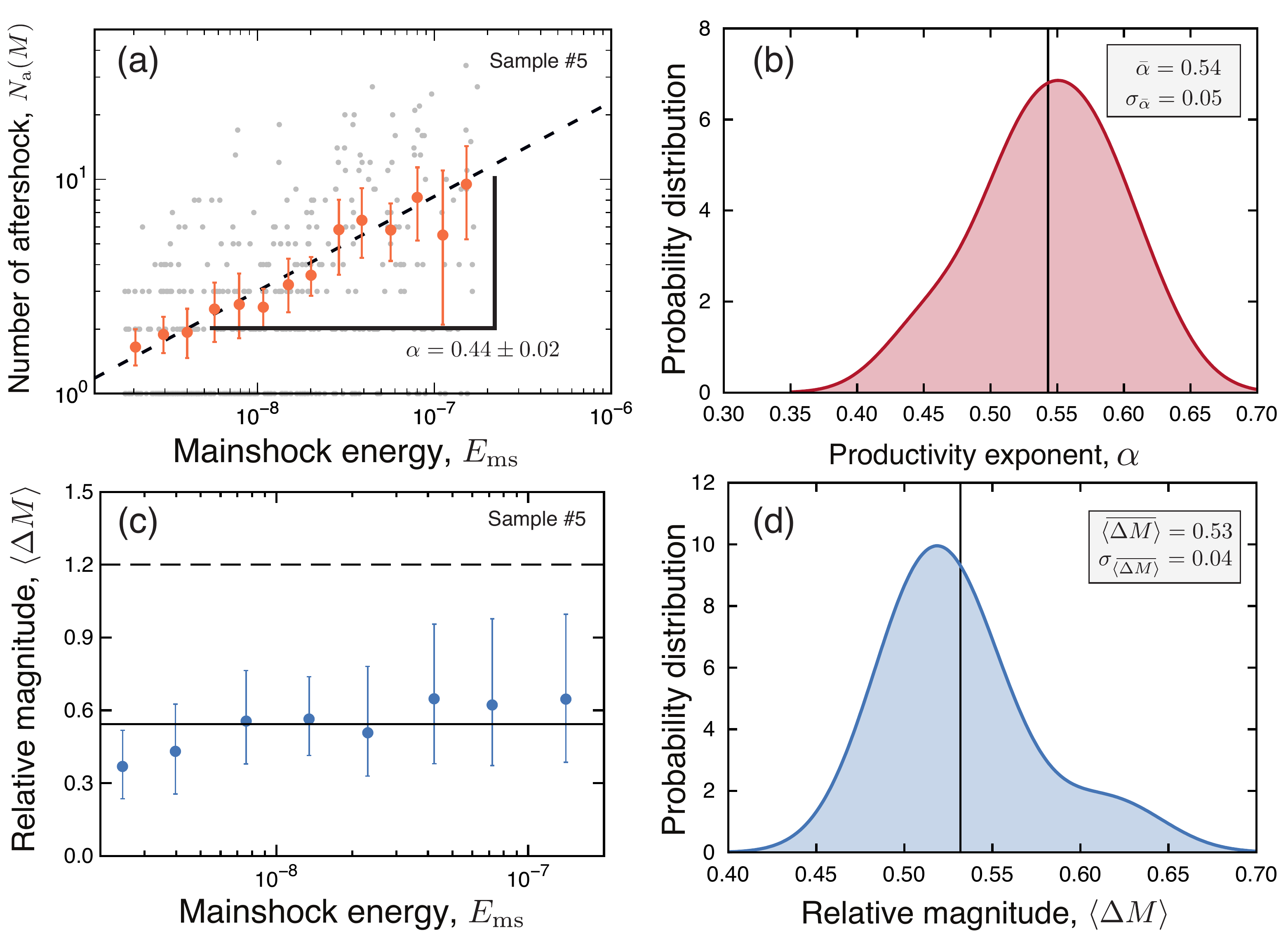}
\caption{The productivity law and the B{\aa}th's law. (a) The relationship between the number of aftershocks $N_{\text{a}}$ and the energy $E_{\text{ms}}$ of the triggering mainshock on log-log scale. Each gray dot represents a sequence of aftershocks (as defined in the Omori's law) occurring one a particular sample (sample \#5) and the red dots are window averages values (error bars stand for $95\%$ bootstrap confidence intervals). The dashed line shows the adjusted power-law relationship, $N_{a}(E_{\text{ms}})\sim E_{\text{ms}}^\alpha$, with $\alpha=0.44\pm0.02$. (b) Probability distribution of $\alpha$ over all samples; average (indicated by a vertical line) and standard deviation of $\alpha$ are shown in the plot. (c) Average difference in magnitude $\langle\Delta M \rangle$ between the mainshock and its largest aftershock as a function of the mainshock energy $E_{\text{ms}}$ (on lin-log scale). The blue dots are averages over all sequences in a given energy windows for one sample (sample \#5) and the error bars are $95\%$ bootstrap confidence intervals. The dashed line indicates the usual B{\aa}th's law predictions for earthquakes ($\langle\Delta M \rangle=1.2$) and the continuous line is the adjusted plateau ($\langle\Delta M \rangle=0.54\pm0.05$). (d) Probability distribution of $\langle\Delta M\rangle$ over all samples; average (indicated by a vertical line) and standard deviation of $\langle\Delta M$ are shown in the plot. All probability distributions are kernel density estimation with bandwidth given by Scott's rule.
}
\label{fig:5}
\end{figure}

\textit{Productivity Law.} Still analyzing the sequence of aftershocks, we quantify the productivity law that measures how the number of aftershocks ($N_{\text{a}}$) triggered by a mainshock increases with mainshock energy ($E_{\text{ms}}$). A power-law relationship, $N_{\text{a}}\sim E_{\text{ms}}^{\alpha}$, with $\alpha$ ranging from $0.7$ to $0.9$ was observed for earthquakes~\cite{Helmstetter_intro}, and also reported for material fractures. For instance, the same relationship with different exponents was found for creep in ice single crystals~\cite{Weiss} ($\alpha\approx0.6$), porous material under compression~\cite{Baro} ($\alpha\approx0.3$) and charcoal under internal stresses~\cite{Ribeiro} (where two power-law regimes appear: $\alpha\approx0.3$ for small $E_{\text{ms}}$ and $\alpha\approx0.8$ for large $E_{\text{ms}}$). By employing the same definitions for mainshocks and aftershocks used in the Omori's law, we count the number of aftershocks $N_{\text{a}}(E_{\text{ms}})$ triggered by a mainshock of energy $E_{\text{ms}}$, in each sample. Figure~\ref{fig:5}(a) shows the relationship between $N_{\text{a}}$ and $E_{\text{ms}}$ on log-log scale for a given sample, where each gray dot is related to an aftershock sequence. Very similar behavior is observed for all samples. Despite the scattering, a statistically significant relationship is observed (Pearson correlation $0.47$, $p$-value~$<10^{-16}$). This relationship becomes even more evident when calculating the window average values of these data (red dots), where the tendency to follow a power law with an exponent $\alpha=0.44\pm0.02$ is clear and holds for around two decades. Figure~\ref{fig:5}(b) shows the probability distribution of $\alpha$ over all samples, where an average value of $\bar{\alpha}=0.54$ and a standard deviation of $\sigma_{\bar{\alpha}}=0.05$ are observed. As in earthquakes and material fractures, this behavior indicates that aftershock triggering results from the stress redistributions.

\textit{B{\aa}th's Law.} Finally but still on the sequence of aftershocks, we focus on the B{\aa}th's law. This empirical law~\cite{Bath} predicts that the average difference in magnitude ($\log E$) between a mainshock and its largest aftershock is independent of the mainshock energy and approximately equal to 1.2. Despite the universal character occasionally ascribed to this empirical law, it is only valid under very strict conditions that are not often observed in real data~\cite{Helmstetter_intro2}. To study this law in our data, we calculate the difference $\Delta M = \log E_{\text{ms}} - \log E_{\text{la}}$ (here $E_{\text{la}}$ is the energy of the largest aftershock in a given sequence) as a function of $E_{\text{ms}}$ for all aftershock sequences (defined as in the Omori's law) obtained from each sample. Figure~\ref{fig:5}(c) shows the average value of this relative difference, $\langle \Delta M \rangle$, as a function of the mainshock energy $E_{\text{ms}}$ for a given sample. The results show that $\langle \Delta M \rangle$ is practically independent of $E_{\text{ms}}$ and well approximated by the constant plateau $\langle \Delta M \rangle=0.54\pm0.05$ for this sample. Figure~\ref{fig:5}(d) shows the probability distribution of this plateau value over all samples, where an average of $\overline{\langle \Delta M\rangle}=0.53$ and standard deviation of $\sigma_{\overline{\langle \Delta M\rangle}}=0.04$ are observed. This average value is much lower than the predictions of the original B{\aa}th's law, which is not surprising under the findings of Helmstetter and Sornette~\cite{Helmstetter_intro2}. By using simulations of the epidemic-type aftershock sequence (ETAS) model, they show that $\langle \Delta M \rangle$ is dependent on both $\beta'$ (of the Gutenberg-Richter law) and $\alpha$ (of the productivity law); roughly speaking, $\langle \Delta M \rangle$ is expect to decrease with $\beta'$ and the shape of $\langle \Delta M \rangle$ versus $E_{\text{ms}}$ is mostly controlled by $\alpha$. They further observe that $\langle \Delta M \rangle$ does not vary significantly with the mainshock energy only for $\alpha<\beta'$, which is in agreement with our results.

\section{Conclusions}
In summary, we reported on an extensive characterization of the acoustic emissions in experiments with crumpled plastic sheets focusing on a comparison with the most fundamental seismic laws. The parallel discussed here fills a hiatus between studies about Gutenberg-Richter law and other seismic laws that had not been reported yet (Omori's, B{\aa}th's, and the productivity law). We verified that these fundamental laws also emerge in the process of uncrumpling plastic sheets. However, the parameters of these laws are (in most cases) different from those observed for earthquakes and fracture experiments, revealing that the acoustic emissions of uncrumpling processes have unique features, which may trigger other investigations with different materials and conditions as well as be useful for comparing models. Finally, we believe that the technical simplicity of these uncrumpling experiments ally with more complete investigations may contribute to a better understating of the mechanisms underlying the complex behavior of systems that emit cracking noises.

\acknowledgments
We are grateful to Capes, CNPq and Funda\c{c}\~ao Arauc\'aria for financial support. HVR acknowledges the financial support of the CNPq under Grant No. 440650/2014-3.


\begin{thebibliography}{0}


\bibitem{Utsu_intro} 
  \Name{Utsu T.}
  \REVIEW{Pure Appl. Geophys.}{155}{1999}{509}.


\bibitem{Godano_intro} 
  \Name{Godano C., Lippiello E., \and de Arcangelis L.}
  \REVIEW{Geophys. J. Int.}{199}{2014}{1765}.

\bibitem{Bak_intro} 
  \Name{Bak P., Christensen K., Danon L., \and Scanlon T.}
  \REVIEW{Phys. Rev. Lett.}{88}{2002}{178501}.

\bibitem{Corral_intro0} 
  \Name{Corral A.}
  \REVIEW{Phys. Rev. E}{68}{2003}{035102(R)}.

\bibitem{Corral_intro} 
  \Name{Corral A.}
  \REVIEW{Phys. Rev. Lett.}{92}{2004}{108501}.


\bibitem{Saichev_intro} 
  \Name{Saichev A. \and Sornette D.}
  \REVIEW{Phys. Rev. Lett.}{97}{2006}{078501}.

\bibitem{Touati} 
  \Name{Touati S., Naylor M., \and Main I. G.}
  \REVIEW{Phys. Rev. Lett.}{102}{2009}{168501}.

\bibitem{Davidsen_intro} 
  \Name{Davidsen J. \and Kwiatek G.}
  \REVIEW{Phys. Rev. Lett.}{110}{2013}{068501}.

\bibitem{Utsu_intro2} 
  \Name{Utsu T., Ogata Y., \and Matsu\'ura S.}
  \REVIEW{J. Phys. Earth}{43}{1995}{1}.



\bibitem{Helmstetter_intro} 
  \Name{Helmstetter A.}
  \REVIEW{Phys. Rev. Lett.}{91}{2003}{058501}.

\bibitem{Bath} 
 \Name{B{\aa}th M.}
 \REVIEW{Tectonophysics}{2}{1965}{483}.

\bibitem{Helmstetter_intro2} 
  \Name{Helmstetter A. \and Sornette D.}
  \REVIEW{Geophys. Res. Lett.}{30}{2003}{2069}.












\bibitem{Hirata} 
  \Name{Hirata T.}
  \REVIEW{J. Geophys. Res.}{92}{1987}{6215}.

\bibitem{Diodati} 
  \Name{Diodati P., Marchesoni F., \and Piazza S.}
  \REVIEW{Phys. Rev. Lett.}{67}{1991}{6215}.

\bibitem{Weiss} 
  \Name{Weiss J. \and Miguel M. C.} 
  \REVIEW{Mater. Sci. Eng. A}{387-389}{2004}{292}.

\bibitem{Davidsen} 
  \Name{Davidsen J., Stanchits S., \and Dresen G.}
  \REVIEW{Phys. Rev. Lett.}{98}{2007}{125502}.

\bibitem{Kun} 
  \Name{Kun F., Costa M. H., Costa-Filho R. N., Andrade Jr J. S., Soares J. B., Zapperi S., \and Herrman H. J.}
  \REVIEW{J. Stat. Mech.}{}{2007}{P02003}.

\bibitem{Kun2} 
  \Name{Kun F., Hal\'asz Z., Andrade Jr J. S., \and Herrmann H. J.}
  \REVIEW{J. Stat. Mech.}{}{2009}{P01021}.

\bibitem{Niccolini} 
  \Name{Niccolini G., Bosia F., Carpinteri A., Lacidogna G., Manuello A., \and Pugno N.}
  \REVIEW{Phys. Rev. E}{80}{2009}{026101}.

\bibitem{Niccolini2} 
  \Name{Niccolini G., Schiavi A., Tarizzo P., Carpinteri A., Lacidogna G., \and Manuello A.}
  \REVIEW{Phys. Rev. E}{82}{2010}{046115}.

\bibitem{Niccolini3} 
  \Name{Niccolini G., Carpinteri A., Lacidogna G., \and Manuello A.}
  \REVIEW{Phys. Rev. Lett.}{106}{2011}{108503}.


\bibitem{Salje} 
  \Name{Salje E. K. H., Lampronti G. I., Soto-Parra D. E., Bar\'o J., Planes A., \and Vives E.}
  \REVIEW{Am. Mineral.}{98}{2013}{609}.

\bibitem{Baro} 
  \Name{Bar\'o J., Corral A., Illa X., Planes A., Salje E. K. H., Schranz W., Soto-Parra D. E., \and Vives E.}
  \REVIEW{Phys. Rev. Lett.}{110}{2013}{088702}.

\bibitem{Nataf} 
  \Name{Nataf G. F., Castillo-Villa P. O., Sellappan P., Kriven W. M., Vives E., Planes A., \and Salje E. K. H.}
  \REVIEW{J. Phys.: Condens. Matter}{26}{2014}{275401}.

\bibitem{Nataf2} 
  \Name{Nataf G. F., Castillo-Villa P. O., Bar\'o J., Illa X., Vives E., Planes A., \and Salje E. K. H.}
  \REVIEW{Phys. Rev. E}{90}{2014}{022405}.

\bibitem{Makinen} 
 \Name{M\"akinen T., Miksic A., Ovaska M., \and Alava M. J.}
 \REVIEW{Phys. Rev. Lett.}{115}{2015}{055501}.

\bibitem{Ribeiro}
  \Name{Ribeiro H. V., Costa L. S., Alves L. G. A., Santoro P. A., Picoli Jr S., Lenzi E. K., \and Mendes R. S.}
  \REVIEW{Phys. Rev. Lett.}{115}{2015}{025503}.

\bibitem{Tsai2} 
 \Name{Tsai S.-T., Wang L.-M., Huang P., Yang Z., Chang C.-D., \and Hong T.-M.}
 \REVIEW{Phys. Rev. Lett.}{116}{2016}{035501}

\bibitem{Sethna2} 
  \Name{Sethna J. P., Dahmen K. A., \and Myers C. R.}
  \REVIEW{Nature}{410}{2001}{242}.


\bibitem{Witten} 
  \Name{Witten T. A.}
  \REVIEW{Rev. Mod. Phys.}{79}{2007}{643}.

\bibitem{Marder} 
  \Name{Marder M., Deegan R. D., \and Sharon E.}
  \REVIEW{Physics Today}{60}{2007}{33}.


\bibitem{Kramer} 
  \Name{Kramer E. M. \and Lobkovsky A. E.}
  \REVIEW{Phys. Rev. E}{53}{1996}{1465}.

\bibitem{Mendes} 
  \Name{Mendes R. S., Malacarne L. C., Santos R. P. B., Ribeiro H. V., \and Picoli Jr S.}
  \REVIEW{EPL}{92}{2010}{29001}.

\bibitem{Houle} 
  \Name{Houle P. A. \and Sethna J. P.} 
  \REVIEW{Phys. Rev. E}{54}{1996}{278}.

\bibitem{Salminen} 
  \Name{Salminen L. I., Tolvanen A. I., \and Alava M. J.}
  \REVIEW{Phys. Rev. Lett.}{89}{2002}{185503}.

\bibitem{Koivisto} 
  \Name{Koivisto J., Rosti J., \and Alava M. J.}
  \REVIEW{Phys. Rev. Lett.}{99}{2007}{145504}.

\bibitem{Tsai} 
 \Name{Tsai S.-T., Chang C.-D., Chang C.-H., Tsai M.-X., Hsu N.-J., \and Hong T.-M.}
 \REVIEW{Phys. Rev. E}{92}{2015}{062925}.


\bibitem{Clauset} 
 \Name{Clauset A., Shalizi C. R., \and Newman M. E. J.}
 \REVIEW{SIAM Rev.}{51}{2009}{661}.
 
%
%
%

\end{thebibliography}
\end{document}